\documentclass[twocolumn,prb,amsmath,amssymb,showpacs]{revtex4}
\usepackage{graphicx}
\usepackage{subfigure}
\pdfoutput=1

\begin{document}
\newcommand{\al}{\alpha}
\newcommand{\be}{\beta}
\newcommand{\de}{\delta}
\newcommand{\D}{\Delta}
\newcommand{\e}{\epsilon}
\newcommand{\s}{\sigma}
\newcommand{\del}{\nabla}
\newcommand{\p}{\partial}
\newcommand{\mbp}{\mathbf{p}}
\newcommand{\mbP}{\mathbf{P}}
\newcommand{\mbq}{\mathbf{q}}
\newcommand{\mbk}{\mathbf{k}}
\newcommand{\mbr}{\mathbf{r}}
\newcommand{\mbm}{\mathbf{m}}
\newcommand{\bs}[1]{\boldsymbol{#1}}
\newcommand{\mbf}[1]{\mathbf{#1}}
\newcommand{\bfhat}[1]{\mathbf{\hat{#1}}}
\newcommand{\mcal}[1]{\mathcal{#1}}
\newcommand{\nn}{\nonumber\\}
\newcommand{\up}{\uparrow}
\newcommand{\bea}{\begin{array}}
\newcommand{\eea}{\end{array}}
\newcommand{\ben}{\begin{equation}}
\newcommand{\een}{\end{equation}}
\newcommand{\bed}{\begin{displaymath} }
\newcommand{\eed}{\end{displaymath}}
\newcommand{\vecprod}[2]{\mathbf{#1}\cdot\mathbf{#2}}
\newcommand{\inftyint}{\int_{-\infty}^{\infty}}
\title {Dissipative dynamics of magnetic solitons in metals}
\author{Clement H. Wong}
\author{Yaroslav Tserkovnyak}
\affiliation{Department of Physics and Astronomy, University of California, Los Angeles, California 90095, USA}

\begin{abstract}
Soliton dynamics in spin-textured metals generate electrical currents, which produce backaction through spin torques.  We modify the Landau-Lifshitz-Gilbert equation and the corresponding solitonic equations of motion to include such higher-order texture effects.   We also find a quasistatic equation for the induced electrochemical potential, which needs to be solved for self-consistently, in the incompressible limit.  As an example, we consider the gyration of a vortex in a point-contact spin valve, and discuss modifications of orbit radius, frequency, and dissipation power.
\end{abstract}



\maketitle
The current-driven dynamics of magnetic solitons such as domain walls, spirals, and vortices has recently attracted much attention.\cite{tataraPRP08,ralphJMMM08} It is actively researched for potential applications in magnetic random-access memory,\cite{yamadaNATM07} data storage devices,\cite{parkinUSP06} and high-quality tunable microwave emitters.\cite{pribiagNATP07}  Soliton dynamics might also be relevant for current-induced switching in single-domain magnets,  where magnetic textures are often nucleated in intermediate stages during the switching process. Spin-valve structures provide an experimentally accessible arena for studying the current or magnetic-field driven dynamics of solitons or multi-solitonic states.\cite{buchananNATP05} It is therefore desirable to have a precise theoretical understanding of the magnetization dynamics in spin-textured metallic magnets.

In the literature, current-driven magnetization dynamics are usually described by the Landau-Lifshitz-Gilbert (LLG) equation with spin torques, or the corresponding equations of motion for collective coordinates describing solitons.  The electrical currents are conventionally treated as an external source and magnetic spin texture is only taken into account through the exchange energy and the associated effective field.  However, it is now established that magnetic texture dynamics generate charge currents,\cite{yangPRL09} which produce feedback on the magnetization via spin torques.\cite{forosPRB08} The three-dimensional coupled spin-charge hydrodynamic equations, including dissipation and thermal noise, were derived in Ref.~[\onlinecite {wongPRB09}].

The purpose of this paper is to theoretically examine the role of such nonequilibrium spin-torque effects on the dynamics of the nanoscale spin-texture solitons. We find, in particular, that the magnitude of the corrections to the magnetic damping can be significant compared to the usual Gilbert damping in conventional transition-metal based materials. The additional terms in the LLG equation are nonlocal and anisotropic and cannot in general be absorbed into existing parameters. In this regard,  they fundamentally modify the dynamics.  Furthermore, because they are sensitive to the details of the magnetic order, they may be used to probe the local geometry and global topology of the magnetic texture.


The  LLG equation governing long-wavelength spin-texture dynamics in metals is given by\cite{tserkovJMMM08}
\ben
S(\dot{\mbf{m}}+\al\mbf{m} \times\dot{\mbf{m}} ) +\mbf{m} \times \mbf{H} = \mcal{P}(1+\beta \mbf{m} \times) ( \mbf{j} \cdot\bs{\del}) \mbf{m}\,,
\label{LLG}
\een
where we denote the local orientation of the spin-density field by the unit-vector field $\mbf{m}(\mbf{r},t)$, $S$ is the magnitude of the saturated spin density, and $\mbf{j}$ is the charge-current density. The magnetization vector is given by  $\mathbf{M}=\gamma S \mathbf{m}$. In terms of the electron charge $-e$, $\mcal{P}\equiv {p\hbar/2e}$ is the charge-to-spin conversion factor, where $p$ is a dimensionless material-dependent spin-polarization parameter, which is typically a number of order 1. Here, $ \mbf{H}$ stands for the effective field defined by the functional derivative $\mbf{H}\equiv \delta_\mbf{m}\mathcal{V} $, where $\mcal{V}[\mbf{m}]$ is the free energy of the ferromagnet. In a planar geometry, the in-plane current $\mathbf{j}$ in Eq.~(\ref{LLG}) has to be solved for self-consistently, taking into account spin-texture generated electromotive forces.\cite{barnesPRL07, {wongPRB09}} Any additional spin torques injected vertically by externally polarized spin currents will be included as contributions $\mbf{H}_ {\rm st} $ to the effective field. Equation \eqref{LLG} has been used to analyze the current-driven dynamics of magnetic vortices, spirals, and domain walls. \cite{thomasNAT06}

To illustrate the essential physics in the simplest setting, we start by considering the highly-compressible limit, such that the spin-charge dynamics may be decoupled, resulting in a modified LLG equation (with the opposite limit discussed later):\cite{wongPRB09,foot0a}
 \ben
\left[(1+\mbf{m}\times\tensor{K})+\mbf{m}\times(\al+\tensor{\Gamma})\right] \dot{\mbf{m}}+\mbf{m}\times\mbf{H}/S=0\,.
\label{LLG2}
 \een
The current-texture interaction generated a correction to Gilbert damping given by
\[
\tensor{\Gamma}= {\mcal{P}^2\over \rho S} \left[( \mbf{m}\times\del_i\mbf{m})\otimes( \mbf{m}\times\del_i\mbf{m})
-{\beta^2} \del_i\mbf{m} \otimes\del_i \mbf{m}\right]\,,
\]
as well as a renormalization of the gyromagnetic tensor:
\[
\tensor{K}= {\beta\mcal{P}^2\over \rho S}\left[(\mbf{m}\times\del_i\mbf{m})\otimes\del_i\mbf{m}-\del_i\mbf{m}\otimes(\mbf{m}\times\del_i\mbf{m})\right]\,.
\label{tensorK}
\] 
Here, $\rho$ is the resistivity, the summation of the repeated spatial indices ($i=x,y,z$) is implied, and both tensors are given to quadratic order in texture gradients.. The total damping tensor determines the dissipation power:
\ben
P\equiv-\int d^3r\,\vecprod{H}{\dot{m}}=S \int d^3r\,\dot{\mbf{m}}\cdot(\al+\tensor{\Gamma})\cdot\dot{\mbf{m}}\,.
\label{diss}
\een
Let us estimate  the size of the correction to the usual Gilbert damping using typical values for transition metals:  resistivity $\rho\sim100~\Omega$\,nm, damping parameters $\al,\beta\sim10^{-2}$, gyromagnetic ratio $\gamma \approx-1.8\times10^{11}$~rad/s\,T, magnetization $M\sim10^6$~A/m, exchange constant $A_{\rm xc}\sim10^{-11}$~J/m, exchange length $\lambda\equiv\sqrt{A_{\rm xc}/\mu_0 M^2}\sim3$~nm, and  polarization factor  $p\sim1$.  We find that numerically, $|\tensor{\Gamma}|\sim{\gamma\mcal{P}^2/\rho M \lambda^2}\sim\al$.  Therefore, for textures with length scales on the order of $\lambda$, the nonlocal damping correction should be treated on equal footing with the usual Gilbert damping, the precise magnitude of the former depending on the material and texture.

Numerical considerations aside, the nonlocal texture effects give the LLG equation qualitatively different structure.  Solving Eq.~\eqref{LLG2} directly is generally a formidable task that requires numerical integration. However, one is often interested in the dynamics of magnetic solitons which are particle-like objects whose motion may be captured by collective coordinates.  One can then approximate Eq.~\eqref{LLG} by an equation of motion for these collective coordinates.\cite{thielePRL73,tretiakovPRL08} In the following, we add the nonlocal texture effects to the latter. We thus consider magnetic textures $\mbf{m}(\mbf{q}(t),\mbf{r})$ whose dynamics are parametrized by generalized coordinates $\mbf{q}(t)$.  For example, they may be the location of topological defects such as vortices and domain walls in rectangular or curvilinear coordinates, or the out-of-plane angles and widths of the N{\'e}el domain walls, or the direction and pitch of spirals. The resulting equations of motion for the generalized coordinates will be generally nonlinear.

Denoting the spherical angles of the spin density field $\mbf{m}(\mbf{q},\mbf{r})$ by $(\theta,\varphi)$, a general magnetic texture is characterized by the following second-rank tensor densities:
\begin{align}
{b}_{ij} &\equiv\mbf{m}\cdot(\p_{q_i}\mbf{m}\times\p_{q_j}\mbf{m})=\sin\theta  (\p_{q_i}\theta\p_{q_j}\varphi-\p_{q_j}\theta \p_{q_i}\varphi)\,,\nn
{d}_{ij}&\equiv\p_{q_i}\mbf{m}\cdot\p_{q_j}\mbf{m}= \p_{q_i}\theta\p_{q_j}\theta+\sin^2\theta\, \p_{q_i}\varphi\p_{q_j}\varphi\,,
\label{TensorD}
 \end {align}  
Since $\hat{b}$ is the Jacobian $\p(-\cos\theta, \varphi )/\p(q_i,q_j)$ of the mapping $(q_i,q_j)\to(-\cos\theta,\varphi)$, if the mapping is nonsingular, its integral $G_{ij}=4\pi Q_{ij}$ can be interpreted in terms of the number of times $Q_{ij}$ the magnetization maps the $(q_i,q_j)$ plane onto the sphere. If the boundary conditions dictate $Q_{ij}$ to take discrete (e.g., integer) values, it becomes a topological invariant, which characterizes topological sectors of the spin texture.\cite{belavinJETPL75} If we take the partial derivatives in Eqs.~(\ref{TensorD}) with respect to the ordinary spatial coordinates $\mbf{r}$, $b_{ij}/4\pi$ is called the topological or skyrmion charge density and $Q_{ij}$ the topological charge. Geometrically, $b_{ij}$ are the oriented surface elements spanned by the gradients of the mapping from the collective coordinates to the magnetization unit sphere, $\mbf{q}\to\mbf{m}$. The symmetric tensor $\hat{d}$ determines the rate at which the local magnetization changes as a function of the generalized velocities, by $\dot{\mbf{m}}^2 =d_{ij}\dot{q}_i\dot{q}_j$, which is proportional to the usual Gilbert dissipation power density.  


Consider first the case without damping, i.e., $\al=\beta=0$ in Eq.~\eqref{LLG}.  This Landau-Lifshitz equation with reactive spin torque follows from the variational principle, $\mbf{m}\times\delta_\mbf{m}S=0$, with a magnetization action $S[\mbf{m}]=\int dt d^3r\mcal{L}$.\cite{tataraPRP08} If instead of varying over the space of all possible magnetization fields, we restrict our variation to the subspace parametrized by the generalized coordinates $\mbf{q}$, constraining the dynamics accordingly, $\dot{\mbf{m}}=\dot{{q}}_i\p_{q_i}\mbf{m}$, we find an equation of motion for $\mbf{q}$ given by $ \int d^3r\p_{q_i}\mbf{m}\cdot(\mbf{m}\times[\rm Eq.~(\ref{LLG})])$. We use the same equation to derive the equation of motion for the generalized coordinates from Eq.~\eqref{LLG2}, including the dissipative terms:  
\ben
[\hat{G}(\mbf{q})+\hat{G}'(\mbf{q})] \dot{\mbf{q}}+[\hat{D}(\mbf{q})+\hat{D}'(\mbf{q})]\dot{\mbf{q}}=\mbf{F}(\mbf{q})\,,
\label{EOM}
\een
where the well-known LLG tensors are given by
\begin{align}
{F}_i&\equiv-{1\over S}\int d^3r \, \p_{q_i}\mbf{m}\cdot \mbf{H}_{\rm st}-{1\over S}\p_{q_i} \int d^3r \, \mathcal{V} \,,\nn
\hat{G}&\equiv \int d^3r\, \hat{b}\,,\,\,\,\hat{D}\equiv\al\int d^3r\, \hat{d}\,.
\label{tensors}
\end{align}
The spin torque corrections to the gyrotropic and damping tensor are given by  ${G}'_ {ij} (\mbf{q})\equiv\int d^3r\p_{q_i}{\mbf{m}}\cdot\tensor{K}\cdot\p_{q_j}{\mbf{m}}$, $D'_{ij} (\mbf{q})\equiv \int d^3r\p_{q_i}{\mbf{m}}\cdot\tensor{\Gamma}\cdot\p_{q_j}{\mbf{m}}$. The additional dissipation power $P'$ from the correction to the damping tensor may be expressed as $P'=D'_{ij}\dot{q}_i\dot{q}_j$. The validity of Eq.~\eqref{EOM} depends on the accuracy with which one parametrizes the dynamical magnetic texture with collective coordinates.  Typically, there will be a hierarchy of hard and soft modes, determined by their respective relaxation times.\cite{tretiakovPRL08}  The softest mode is usually the rigid translation of the entire texture.  For simplicity, this is the only degree of freedom we will retain, letting $\mbf{R}$ denote the position of the defect center.

Within each topological sector, solitonic configurations minimize the free energy $U=\int d^3r\,{\mathcal{V}}/S$, which includes both exchange and nonlocal magnetostatic interactions sensitive to the shape of the ferromagnet.  For simplicity, we assume that the size of solitons are much smaller than the system size, neglecting boundary effects, so that the internal energy does not depend on the position of the solitons.  The only role that the internal energy will play is to determine the profile ${\mbf{m}}({\mbf{r}}-\mbf{R}(t))$, which is assumed undistorted under spatial translations. Therefore, in the evaluation of the tensors in Eq.~\eqref{TensorD}, we may substitute $\p_\mbf{R}=-\p_\mbf{r}$. In the following, when evaluating the tensors \eqref{TensorD}, we will thus take the derivatives with respect to spatial coordinates: $\p_{q_i}\equiv\p_{R_i}=-\p_{r_i}\equiv-\del_i$. The spin-torque corrections to Eq.~\eqref{tensors} thus become:
\ben
\{\hat{G}',\hat{D}'\}=-{\mcal{P}^2\over\rho S}\int d^3r\{\beta(\hat{b}\hat{d}+ \hat{d}\hat{b}),(\hat{b}^2+\beta^2\hat{d}^2)\}\,.
\een

To express Eq.~\eqref{EOM} in a vector form, we define the gyrovector $\mbf{G}$ by $G_i=\e^{ijk}{G}_{jk}/2$ (along with the analogous definition for $\mbf{b}$ in terms of $\hat{b}$):
\ben
\dot{\mbf{R}}\times(\mbf{G}+\mbf{G}')+(\hat{D}+\hat{D}')\dot{\mbf{R}}=-\p_\mbf{R}U +\mbf{F}^{\rm st}(\mbf{R})\,,
\label{vectorEOM}
\een
where $F^{\rm st}_i \equiv\int d^3r\,\del_{i}\mbf{m}\cdot \mbf{H}_ {\rm st}/S$.  Equation \eqref{vectorEOM} can be thought as describing a massless particle moving with friction in a magnetic field given by $\mbf{G}$ (which is sometimes called the Magnus force), and an effective potential $U(\mbf{R})$ that includes exchange, anisotropy, and magnetostatic energies, driven by an external force $\mbf{F}^{\rm st}$, which may or may not be conservative.  


As an example, we consider two-dimensional (2D) vortex dynamics on a circular disk of radius $L$ of negligible thickness.  The tensors below are to be understood as per unit length in thickness.   We will denote the position $\mbf{r}$ with respect to the center of the disk with polar coordinates $(r,\phi)$, with the vortex center being at $(R,\psi)$. The vortex spin texture is described by
\ben
\varphi(\mbf{r}')=\pi/2+\phi'\,,\,\,\,\theta(\mbf{r}')=2\arctan(r'/\lambda)\,\,\,{\rm at}~r'\leq\lambda\,,
\label{profile}
\een
and $\theta(r'>\lambda)=\pi/2$, where $\mbf{r}'\equiv\mbf{r}-\mbf{R}$, whose polar coordinates are $r'=|\mbf{r}-\mbf{R}|$ and $\phi'(\mbf{R})=\arg (\mbf{r}-\mbf{R})$. Such a vortex profile crudely captures a compromise between the exchange and magnetostatic energies.\cite{guslienkoJAP02} In the coordinate system centered on the vortex, the texture gradients are given by $\bs{\del}\theta=\bfhat{r}\p_r\theta$, $\bs{\del}\varphi=\bs{\hat{\phi}}/r$, so that $d_{rr}= (\p_r\theta)^2$, $d_{\phi\phi}= (\sin\theta/ r)^2$, $d_{r\phi}=0$, and $b_{r\phi}= -{(\p_r\cos\theta)/r}$. The gyrovector density is $\mbf{b}=b_{r\phi}\bfhat{z}$, and the damping tensor density is $\hat{d}= d_{rr}\bfhat{r}\otimes\bfhat{r}+d_{\phi\phi}\bs{\hat{\phi}}\otimes\bs{\hat{\phi}}$. Our vortex profile has the property that $\p_r\theta= \sin\theta/r=f(x)/\lambda$ inside the core, where $x=r/\lambda$ and $f(x)=2/(1+x^2)$, so that
\begin{align}
b_{r\phi}&=d_{rr}=d_{\phi\phi}=(f/\lambda)^2\,\,\,{\rm at}\,\,\,r<\lambda\,;\nonumber\\
b_{r\phi}&=d_{rr}=0\,,\,\,\, d_{\phi\phi}={1/(\lambda x)^2}\,\,\,{\rm at}\,\,\,r>\lambda\,.
\label{b}
\end{align}
In 2D, the gyrovector has only one nonzero component perpendicular to the plane, $\mbf{G}=G\mbf{\hat{z}}$, where $G=G_{r\phi}$.  Up to continuous deformations, it depends only on the vortex profile in the core, where our vortex is a soliton with  topological charge $Q=1/2$, so that $G=4\pi Q=2\pi$,  as one can verify explicitly. Computing the spin-torque correction for the vortex profile \eqref{profile}, we find
\ben
G'= G'_ {r\phi}
=-{\beta\mcal{P}^2\over\rho S \lambda^2}  \int_0^1 dx \,  4\pi x \left({2\over1+x^2}\right)^4\,.
\een
The integral evaluates to about $30$. For the previously listed parameters, we find this $O(\beta)$ correction to be small, $|G'|/G\sim10^{-4}$, which can be safely neglected.  Consider the damping tensor $\hat{D}$ in Eq.~\eqref{tensors}.  Because we assume a rigid vortex which has rotational symmetry about its core, this tensor is diagonal: $\hat{D}=D={\rm Tr}[\hat{D}]/2= \al\int d^2r \, (\nabla_i\mbf{m})^2/2$, which is proportional to the total exchange energy. Recalling that the vortex is far away from any boundaries, we find $D=\al[2\pi+\pi\ln(L/\lambda)]$. The $2\pi$ comes from the region inside the core, where the vortex \eqref{profile} satisfies the minimum energy condition $U_{\rm xc}/A_{\rm xc}=D/\al=G$.  The logarithmic factor comes from  outside the core and is generic to any 2D vortex.  For the spin-torque correction to damping, we neglect the $O(\beta^2)$ contribution outside the core, so that
\ben
 D'
 ={\mcal{P}^2\over\rho S \lambda^2}\int_0^1 dx \,  2\pi x \left({2\over1+x^2}\right)^4\,.
\label{D'}
\een 
Taking disk of radius $L\sim10^3\lambda$, we find that the damping is enhanced by $D'/D\sim\mcal{P}^2/\al\rho S \lambda^2\sim1$.

\begin{figure}[t]
\includegraphics[width=\linewidth]{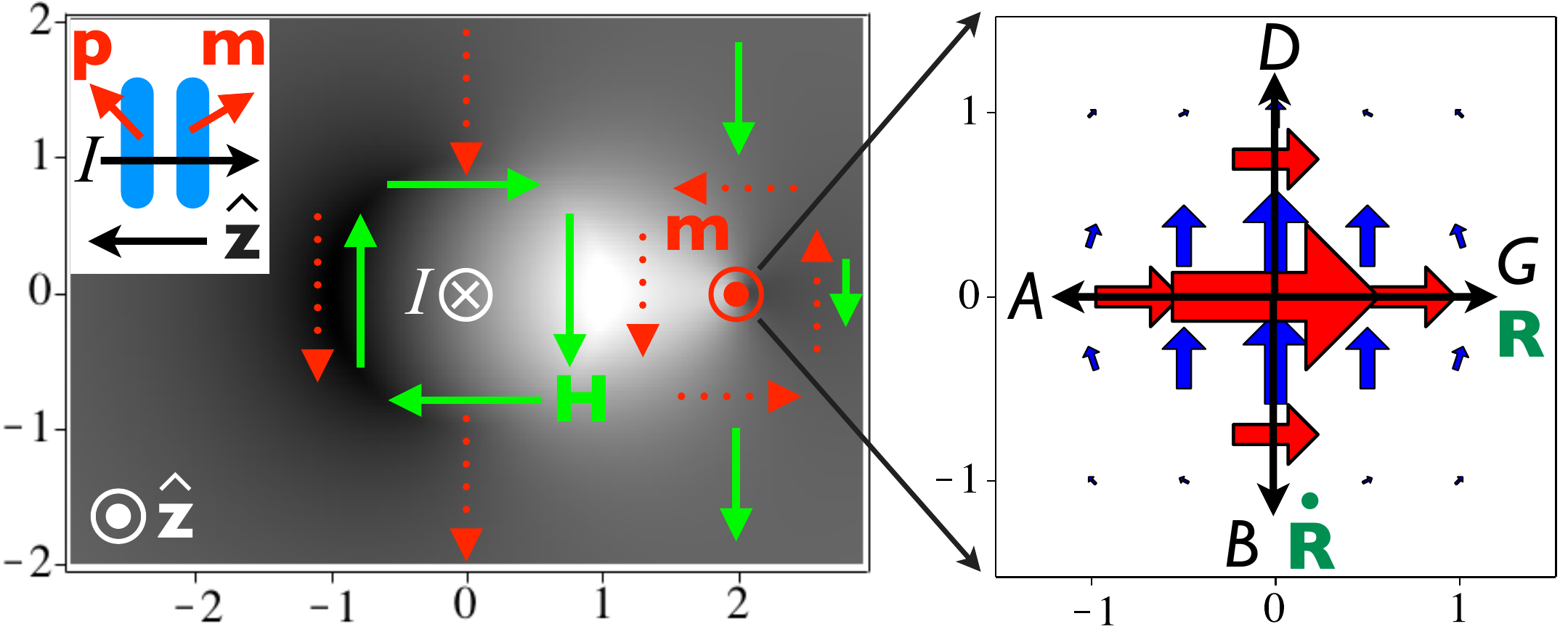}
\caption{(Color online) Left panel: Energy density \eqref{energy} of the vortex texture $\mathbf{m}$ (red dotted arrows) due to the Oersted field $\mathbf{H}$ (green solid arrows), in the $xy$ plane centered at the point contact, with unit length $a$. The arrows depict a crude model of the vortex as four domains separated by $\pi/2$ walls emanating from the vortex center.  On the gray scale, higher energy density corresponds with lighter shade.  One clearly sees a region of high energy that scales linearly with the separation distance between the vortex and the point contact, resulting in a linear attractive potential. The inset shows a side-view schematic of the magnetic bilayer in the $z$ direction. Right panel: A plot of the charge current due to each term in the spin-force field \eqref{sforce} in a region of $3\lambda\times3\lambda$ centered at the vortex core. The contribution collinear with the Magnus force is shown with thick red arrows. The viscous $O(\beta)$ terms are indicated by blue arrows and have been magnified by setting $\beta\sim1$. The long black arrows indicate the direction of each term in Eq.~\eqref{VEOM}, labeled by their respective couplings.}
\label{fig}
\end{figure}

To explore the consequences of our corrections,  consider large-amplitude vortex dynamics in a point contact spin valve, driven by a dc current applied perpendicular to the plane.\cite{mistralPRL08}  Electrons pass from a pinned ferromagnetic layer with magnetization along the direction $\mathbf{p}$, become spin-polarized along the same direction and exert a spin torque on a free layer with a 2D vortex. For simplicity, we assume the free layer is circular and $\mathbf{p}$ is uniform. Consider the potential energy of the vortex, $U (\mathbf{R}) =-\int d^2r\, \mbf{H}\cdot\mbf{M}/S$, due to the Oersted field $\mbf{H}=-h(r)I\bs{\hat{\phi}}$, created by the charge current $-I\bfhat{z}$, where $h(r)$ is the radial profile.  Assuming the current is applied to a circular region of radius $a$,  $h(r)={\mu_0/ 2\pi r} $ at $r>a$.
 The potential energy  (per $S$) is thus given by 
\ben
U(R)=\gamma I\int dr d\phi\, rh(r)\sin\theta(r')\bs{\hat{\phi}}\cdot\bs{\hat{\phi}}'\,,
\label{energy}
\een
where $\bs{\hat{\phi}}'=\mathbf{\hat{z}}\times\mathbf{\hat{r}}'$. When the vortex is outside of the current distribution, the potential is approximately linear in $R$.\cite{mistralPRL08}  See Fig.~\ref{fig}. We will therefore consider the linear potential $U(R)\approx{\rm const}+AIR$, where on dimensional grounds,  $A\sim-\gamma a\mu_0\sim10^{-3}$~m$^3$/sA, taking $a\sim100$~nm.  

In addition to the Oersted field, the current exerts a spin torque on the free-layer magnetization which drives vortex precession.  We take a simple model, assuming that the transverse spin angular momentum of the polarized current is absorbed with an angle-independent efficiency $\s$ by the free layer.  This spin torque can then be included in Eq.~\eqref{LLG2} by adding to the effective field a term $\mbf{H}_{\rm st}=(\s\mcal{P}{I}/\pi a^2)\mbf{p}\times\mathbf{m}$. \cite{ralphJMMM08,slonczewskiJMMM96} Inside the point contact,  $\del_{i}\mbm\approx-\sin\theta\del_i\varphi\mathbf{\hat{r}}'$ (when the vortex core is outside), so that only the out-of-plane component of the polarization  $p_\perp=\mathbf{p}\cdot\mathbf{\hat{z}}$, which creates an in-plane effective field along $\mathbf{\hat{z}}\times\mathbf{m}=-\sin\theta\hat{\mbr}'$, contributes to the resulting force
\ben
\mbf{F}^{\rm st}\approx{\s\mcal{P}p_\perp{I}\over \pi a^2S} \int_{r<a}  d^2r\sin^2\theta\,\boldsymbol{\del}\varphi 
\approx-{\s \mcal{P}p_\perp I\over SR}\boldsymbol{\hat{\psi}}\,,
\een
where $\bs{\hat{\psi}}=\mathbf{\hat{z}}\times\mathbf{\hat{R}}$ and $\mathbf{R}=\mathbf{\hat{R}}R$. Here, we have taken the limit $R\gg a$, approximating the integrand with its value at the origin. Defining $B=\s\mcal{P}p_\perp/S$, Eq.~\eqref{vectorEOM} in the region  $R>a$ becomes
\ben
G\mbf{\hat{z}}\times\dot{\mbf{R}}-D\dot{\mbf{R}}-AI\mbf{\hat{R}}-BI\mbf{\hat{z}}\times\mbf{\hat{R}}/R=0\,,
\label{VEOM}
\een
where we absorb $D'$ in $D$ hereafter. It is natural to switch to complex notation, $Z=X+iY=Re^{i\psi}$, and solve Eq.~\eqref{VEOM} in polar coordinates: 
\ben
\dot{R}+iR\dot{\psi}
={-I}\frac{ (AD-BG/R)+i(AG+BD/R)} {D^2+G^2}\,.
\label{VEOM1}
\een
A circular orbit, $\dot{R}=0$, exists if  $\text{sign}(BG)>0$, which for our vortex ($G>0$) requires $\text{sign}(B)=\text{sign}(p_\perp)>0$.   The orbit has radius $R_0=BG/AD$ and frequency $\dot{\psi}=-AI/GR_0$, and is stable only for $I>0$ corresponding to an attractive potential $U(R)$. A twofold texture enhancement of $D$ would halve the orbit radius and double the frequency and dissipation power, $P=D({AI}/{G})^2$.


We now return to relax the assumption of high compressibility, which underlies derivation of Eq.~\eqref{LLG2}.  The local Ohm's law is given by \cite{wongPRB09}
 \ben
\rho j_i= \mcal{P}(\beta\nabla_i\mbf{m}+\mbf{m}\times\nabla_i\mbf{m})\cdot\dot{\mbf{m}}+\nabla_i\mu/e\,,
\label{current}
\een
 where $\rho$ is the drude resistivity, $\mu(n)=n/K$ is the chemical potential, $n$ is the nonequilibrium particle density and $K$ is the compressibility.     
 In Eq.~\eqref{LLG2}, we decoupled the current-magnetization dynamics by taking $\mu\to0$ and substituting Eq.~\eqref{current} into \eqref{LLG}. For a finite compressibility, however, there is a particle-diffusion current  $\mbf{j}_D=\bs{\del}\mu/e\rho$ that gives an additional spin torque $\bs{\tau}_D$ in Eq.~\eqref{LLG}.\cite{foot5} This spin torque couples the magnetization dynamics to the charge density, which is determined by the continuity equation, $e\dot{n}=\bs{\del}\cdot\mbf{j}$.

We consider a typical scenario in which Coulombic electron-electron repulsion renders electric flows essentially incompressible, and solve these coupled equations in the $n\to0$ limit.  The continuity equation then determines the electrochemical potential according to
\ben
\del^2\mu=\mcal{P}e\boldsymbol{\del}\cdot\mbf{F},
\label{poisson}
\een
where we wrote Eq.~\eqref{current} as $\mbf{j}=-\mcal{P}\mbf{F}/\rho+\mbf{j}_D$.  The texture-dependent field $\mbf{F}=\mbf{e}+\mbf{f}$ (force per unit of ``spin charge" $\mcal{P}e$) is due to the fictitious electric field $\mathbf{e}$ and dissipative field $\mathbf{f}$ with components $e_i= \mbf{m}\cdot(\dot{\mbf{m}}\times\nabla_i\mbf{m})$ and $f_i=-\beta\dot{\mbf{m}}\cdot\nabla_i\mbf{m}$. A rigid soliton in motion generates $\mbf{e}=\dot{\mbf{R}}\times\mbf{b}$ and $\mbf{f}=\beta\hat{d}\cdot\dot{\mbf{R}}$. Formally, Eq.~\eqref{poisson} is Poisson's equation for a charge distribution with polarization $\mbf{F}$, and one can apply standard Green's function methods to solve for $\mu$, once boundary conditions are specified.\cite{yangPRL09} Our vortex generates the force field
\ben
\mbf{F}(\mbf{r}')=\left\{
 \begin{array}{cc}(\beta\dot{\mbf{R}}+\dot{\mbf{R}}\times\bfhat{z})\left(f(x')/\lambda\right)^2\,,& r'\leq\lambda\\ 
\beta(\dot{\mbf{R}}\cdot\bs{\hat{\phi'}})\bs{\hat{\phi'}}/(\lambda x')^2\,,&r'>\lambda
\end {array} . \right.
\label{sforce}
\een
See the right panel of Fig.~\ref{fig}, where the charge current generated by $\mathbf{F}$ is plotted.  The divergence of this current induces a counterbalancing diffusion current $\mathbf{j}_D$, according to Eq.  \eqref{poisson}, which ensures that the total current is divergenceless. Neglecting the $O(\beta)$ terms, the polarization charge $n_{\rm pol}$ is given by 
\ben
n_{\rm pol}(\mbf{r}')\equiv-\frac{\mcal{P}e}{4\pi}\bs{\del}\cdot\mbf{F}=\frac{\mcal{P}e}{4\pi\lambda^3}\bfhat{r}'\cdot(\mathbf{\hat{z}}\times\dot{\mbf{R}})\p_{x'}(f^2)\,.
\een
Since the charges vanish at the boundary, we obtain the Coulombic solution $\mu(\mbf{r})=\int d^3r' n_{\rm pol} (\mbf{r}')/|\mbr-\mbr'| $.

In summary, we have extended the LLG phenomenology to include spin-texture effects stemming from dynamically generated electric currents, and examined these effects in 2D vortex motion under an applied point-contact current.  Our theory can be applied to multi-soliton dynamics, with or without applied currents and magnetic fields, in any dimensions. With a more detailed parametrization of the soliton profile, its equations of motion may be rendered more realistic. The $\beta$-like viscous coupling between current and magnetization dynamics can have giant enhancement\cite{halsPRL09} in dilute magnetic semiconductors due to the nonadiabatic spin-torque effects that are beyond the scope of Eq.~\eqref{LLG}. This should motivate further study in the present context.

We acknowledge support by the Alfred~P. Sloan Foundation and the NSF under Grant No. DMR-0840965.


\begin{thebibliography}{23}
\bibitem{foot5} The resistivity tensor also acquires nonlocal corrections quadratic in texture gradients,\cite{wongPRB09} which would enter in the spin torque at quartic order and does not concern us here.
\end{thebibliography}
\end{document}